\documentclass[]{aastex631}
\usepackage{amsmath}
\usepackage{comment}
\usepackage{textgreek}
\usepackage{multirow}

\shorttitle{}
\shortauthors{Kato et al., 2025}
\graphicspath{{./}{figures/}}

\begin{document}

\title{A discussion on the origin of the sub-PeV Galactic gamma-ray emission}

\author{S. Kato}
\altaffiliation{Corresponding author}
\affiliation{Institut d'Astrophysique de Paris, CNRS UMR 7095, Sorbonne Université, 98 bis bd Arago 75014, Paris, France; sei.kato@iap.fr}
\author{R. Alves Batista}
\affiliation{Institut d'Astrophysique de Paris, CNRS UMR 7095, Sorbonne Université, 98 bis bd Arago 75014, Paris, France; sei.kato@iap.fr}
\author{M. Anzorena}
\affiliation{Institute for Cosmic Ray Research, University of Tokyo, Kashiwa 277-8582, Japan}
\author{K. Awai}
\affiliation{Institute for Cosmic Ray Research, University of Tokyo, Kashiwa 277-8582, Japan}
\author{D. Chen}
\affiliation{National Astronomical Observatories, Chinese Academy of Sciences, Beijing 100012, China}
\author{K. Fujita}
\affiliation{Institute for Cosmic Ray Research, University of Tokyo, Kashiwa 277-8582, Japan}
\author{R. Garcia}
\affiliation{Institute for Cosmic Ray Research, University of Tokyo, Kashiwa 277-8582, Japan}
\author{J. Huang}
\affiliation{Key Laboratory of Particle Astrophysics, Institute of High Energy Physics, Chinese Academy of Sciences, Beijing 100049, China}
\author{G. Imaizumi}
\affiliation{Institute for Cosmic Ray Research, University of Tokyo, Kashiwa 277-8582, Japan}
\author{T. Kawashima}
\affiliation{Institute for Cosmic Ray Research, University of Tokyo, Kashiwa 277-8582, Japan}
\author{K. Kawata}
\affiliation{Institute for Cosmic Ray Research, University of Tokyo, Kashiwa 277-8582, Japan}
\author{A. Mizuno}
\affiliation{Institute for Cosmic Ray Research, University of Tokyo, Kashiwa 277-8582, Japan}
\author{M. Ohnishi}
\affiliation{Institute for Cosmic Ray Research, University of Tokyo, Kashiwa 277-8582, Japan}
\author{C. Pr\'evotat}
\affiliation{Institut d'Astrophysique de Paris, CNRS UMR 7095, Sorbonne Université, 98 bis bd Arago 75014, Paris, France; sei.kato@iap.fr}
\author{T. Sako}
\affiliation{Institute for Cosmic Ray Research, University of Tokyo, Kashiwa 277-8582, Japan}
\author{T. K. Sako}
\affiliation{Nagano Prefectural Institute of Technology, Ueda, 386-1211, Japan}
\author{F. Sugimoto}
\affiliation{Institute for Cosmic Ray Research, University of Tokyo, Kashiwa 277-8582, Japan}
\author{M. Takita}
\affiliation{Institute for Cosmic Ray Research, University of Tokyo, Kashiwa 277-8582, Japan}
\author{Y. Yokoe}
\affiliation{Institute for Cosmic Ray Research, University of Tokyo, Kashiwa 277-8582, Japan}

\begin{abstract}
  Galactic diffuse gamma-ray flux measured by the Tibet AS$\gamma$ experiment and the total Galactic gamma-ray flux measured by Large High Altitude Air Shower Observatory (LHAASO) are found to be consistent, within the statistical and systematic uncertainties, for the inner Galactic Plane region in the sub-PeV energy range ($E>10^{14}\, {\rm eV}$). The result suggests that the sub-PeV Galactic gamma-ray flux is dominated by the diffuse emission. On the other hand, the LHAASO observations suggest that the sub-PeV gamma-ray sources presented in the first LHAASO catalog possibly give a significant contribution to the total sub-PeV Galactic gamma-ray emission ($\approx 60\%$). However, the estimate must be regarded as a conservative upper limit in the sub-PeV energy range. In fact, current gamma-ray observations imply that many of the sub-PeV gamma-ray sources detected by LHAASO have a cutoff or significant softening in their energy spectra in the several tens of TeV energy range, and the resolved-source contribution to the total sub-PeV Galactic gamma-ray emission should be much lower than the above estimate. More sophisticated discussion about the origin of the sub-PeV Galactic gamma-ray emission requires detailed spectral studies of the individual gamma-ray sources and an accurate estimate of the contamination of the source fluxes from the diffuse emission.
\end{abstract}


\section*{}
Accurate measurements of Galactic diffuse gamma-ray emission (GDE) in the sub-PeV energy range ($E > 10^{14}\, {\rm eV}$) are important to study the distribution of PeV cosmic rays ($E > 10^{15}\, {\rm eV}$) and their propagation process in the Galaxy \citep{Lipari_and_Vernetto_PRD_2018, Zhang_2023, Luque_2023, Luque_2025, PhysRevD.110.103035}. Sub-PeV GDE was first detected by the Tibet AS$\gamma$ experiment \citep{TibetDiffuse}, later followed by the Large High Altitude Air Shower Observatory (LHAASO, \citealp{PhysRevLett.131.151001}). Their measurements of the GDE flux in the inner Galactic Plane region ($25^{\circ}<l<100^{\circ}$ and $|b|<5^{\circ}$ for Tibet AS$\gamma$ and $15^{\circ}<l<125^{\circ}$ and $|b|<5^{\circ}$ for LHAASO) differ by a factor of five in the 100 TeV energy range. This apparent discrepancy could, in principle, be attributed to gamma-ray sources not resolved by Tibet AS$\gamma$ \citep{Vecchiotti_2022, Yan_et_al_NatAstron_2024}. Nevertheless, \cite{Kato_2024_2} found that the sub-PeV gamma-ray sources presented in the first LHAASO catalog \citep{1LHAASOCatalog}, the most updated and complete gamma-ray source catalog in the northern hemisphere, only provide a minor contribution to the Tibet GDE. The authors conclude that the difference between the Tibet and LHAASO GDE measurements would come from a large sky-masking scheme adopted by LHAASO; they try to completely remove the contamination by resolved gamma-ray sources, which should in turn lose a significant fraction of the GDE emission along the Galactic Plane. The Tibet GDE flux is likely dominated by GDE from its consistency with the theoretical GDE model given by \cite{Lipari_and_Vernetto_PRD_2018}, while its hadronic nature is supported by the Galactic neutrino observation by IceCube using the $\pi^0$ model template \citep{IceCube_2023, Ackermann_2012} and a data-driven estimate of the leptonic contribution from unresolved gamma-ray sources associated with pulsars \citep{Kaci_2024}. On the other hand, the hadronic nature of the sub-PeV LHAASO GDE flux is supported by \cite{Fang_2023_ApJL, Kaci_2024}. In particular, \cite{Kaci_2024} estimated the contribution from unresolved pulsar-associated gamma-ray sources to be less than $20\%$ of the LHAASO GDE flux above 100 TeV.

\begin{figure}
  \centering
  \includegraphics[scale=0.6]{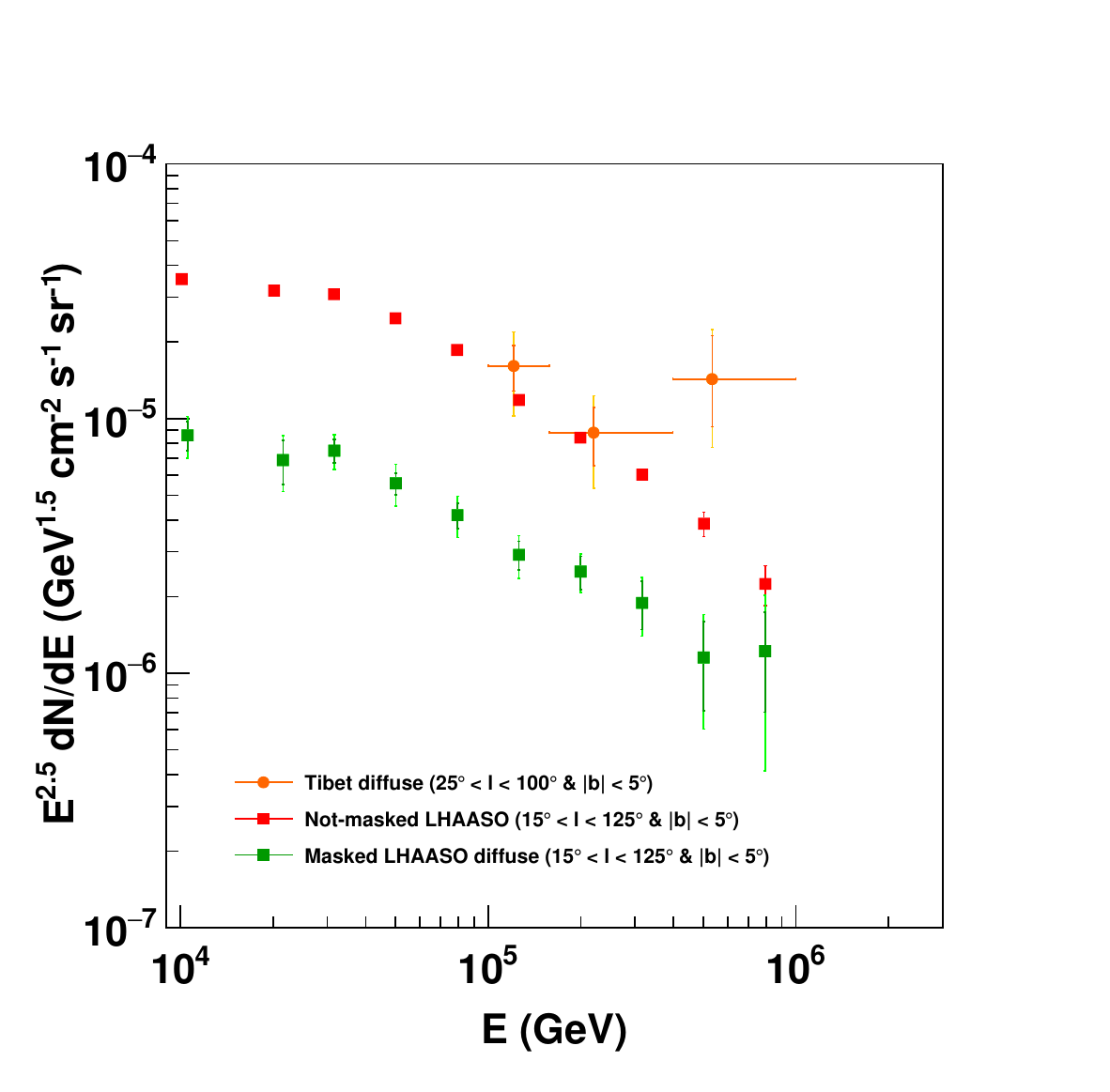}
  \caption{Tibet GDE flux in the sky region of $25^{\circ}<l<100^{\circ}$ and $|b|<5^{\circ}$ (orange, \citealp{TibetDiffuse}) and the not-masked LHAASO flux and the LHAASO GDE flux in $15^{\circ}<l<125^{\circ}$ and $|b|<5^{\circ}$ (red and green, respectively, \citealp{LHAASO_diffuse_WCD_2024}). For the Tibet (LHAASO) GDE flux, the statistical error is shown with the orange (thick-green) vertical bars, while the yellow (light-green) vertical bars show the quadrature sum of the statistical and systematic errors. For the not-masked LHAASO flux, only the statistical error is shown with vertical bars.}
  \label{fig:comparison}
\end{figure}

Recently, the LHAASO Collaboration published the result of their updated GDE measurement extending to the lower energy range of 1 TeV by using the Water Cherenkov Detector Array (WCDA, see the Supplemental Material of \citealp{LHAASO_diffuse_WCD_2024}). Interestingly, they also present the measurements of the gamma-ray energy spectra in the inner ($15^{\circ}<l<125^{\circ}$ and $|b|<5^{\circ}$) and outer Galactic Plane regions ($125^{\circ}<l<235^{\circ}$ and $|b|<5^{\circ}$) without adopting a source-masking scheme; hereafter called the {\it not-masked LHAASO spectra}. These measurements represent the total Galactic gamma-ray emissions from the corresponding sky regions. Figure \ref{fig:comparison} compares the Tibet GDE flux and the not-masked LHAASO flux in the inner Galactic Plane region and shows that their flux levels are consistent within statistical and systematic uncertainties. The Tibet GDE flux at the highest energy is still consistent with the not-masked LHAASO flux within a $\simeq 2 \, \sigma$ statistical deviation. In fact, the Tibet flux point includes ten gamma-ray-like excess events above 398 TeV, and four out of the ten events come from the Cygnus Cocoon region \citep{TibetDiffuse, Fang_2021}. The aforementioned $\simeq 2\, \sigma$ statistical deviation can thus be interpreted as a statistical fluctuation in the observation by Tibet AS$\gamma$. The consistency between the Tibet GDE flux and the not-masked LHAASO flux suggests that the sub-PeV Galactic gamma-ray emission is indeed dominated by GDE.

\begin{figure}
  \centering
  \includegraphics[scale=0.6]{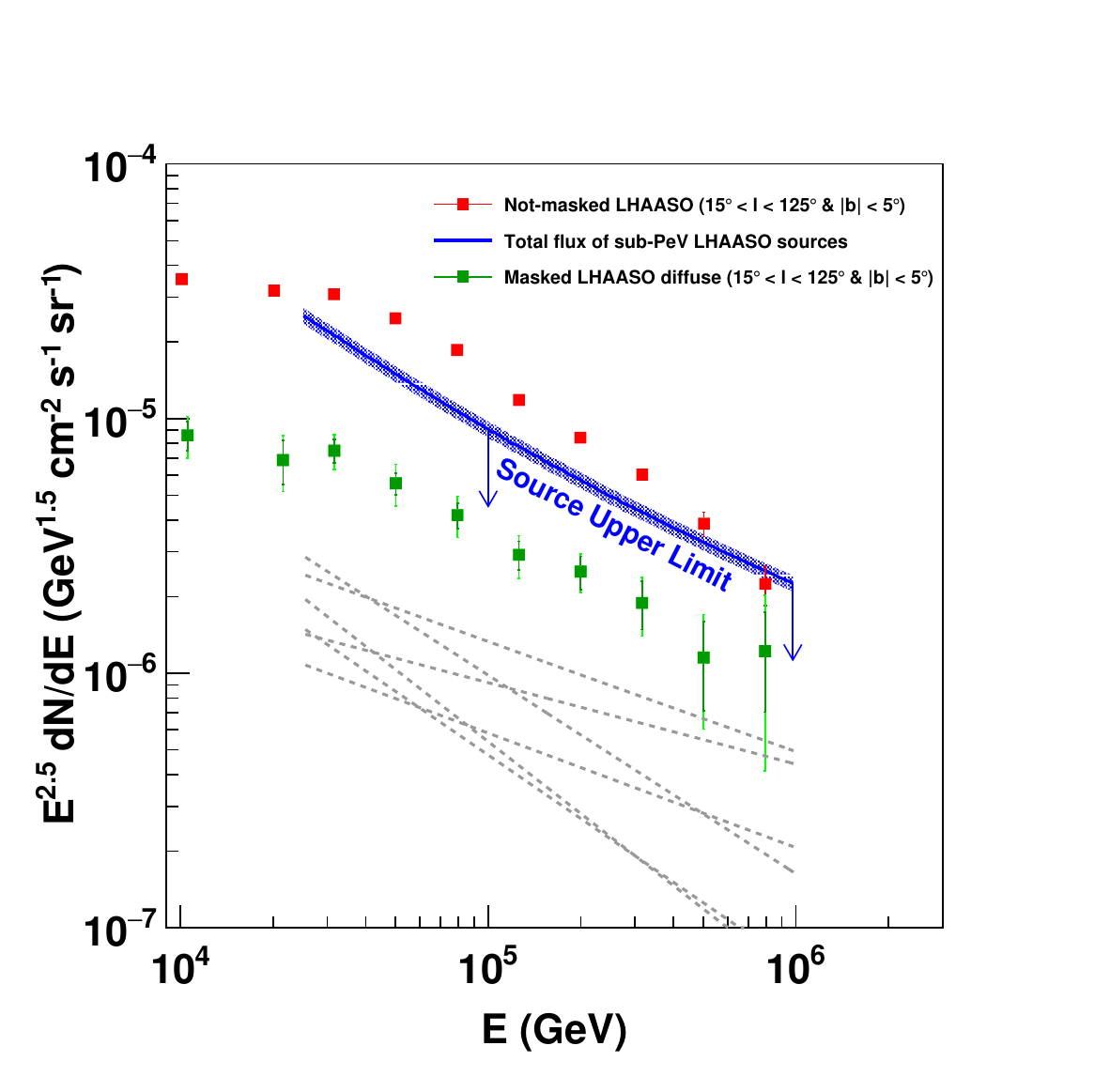}
  \caption{Not-masked LHAASO flux in the inner Galactic Plane region of $15^{\circ}<l<125^{\circ}$ and $|b|<5^{\circ}$ (red) compared with the total flux of the sub-PeV LHAASO sources in the same sky region (blue curve). The blue shaded band indicates the uncertainty of the total source flux. The total source flux must be regarded as a conservative upper limit in the sub-PeV energy range, as emphasized with the notation "Source Upper Limit" and the blue downward arrows put at 100 TeV and 1 PeV; see the text for details. The gray dashed lines are the best-fit power-law spectra of several sub-PeV LHAASO sources above 25 TeV that significantly contribute to the total source flux; see the text. The green points are the same as in Figure \ref{fig:comparison}.}
  \label{fig:nomasked-LHAASO-1LHAASO-sources}
\end{figure}

On the other hand, one can make another possible interpretation. Figure \ref{fig:nomasked-LHAASO-1LHAASO-sources} compares the not-masked LHAASO flux with the total flux of the first LHAASO catalog sources detected above 100 TeV \citep{1LHAASOCatalog} and the Cygnus Cocoon \citep{LHAASOCOLLABORATION2024449} in the inner Galactic Plane region of $15^{\circ}<l<125^{\circ}$ and $|b|<5^{\circ}$. Hereafter, these gamma-ray sources are called the {\it sub-PeV LHAASO sources}. The total flux of the sub-PeV LHAASO sources and its uncertainty is calculated following the methodology presented in \cite{Fang_2023_ApJL, Kato_2024_2} assuming that the energy spectra of the sub-PeV LHAASO sources follow the best-fit power-law spectra given by \cite{1LHAASOCatalog, LHAASOCOLLABORATION2024449}. Note that the above total source flux is a simple sum of all contributions from the sub-PeV LHAASO sources; it is different from the source flux estimated by \cite{Kato_2024_2}, where the authors did not account for the source contribution from the sky regions masked in the Tibet GDE analysis for the comparison with the Tibet GDE flux. The blue shaded band indicates the quadrature sum of the statistical and systematic uncertainties in the measurement by the Kilometer Squared Array (KM2A) of LHAASO. The gray dashed lines are the best-fit power-law spectra of several sub-PeV LHAASO sources above 25 TeV that significantly contribute to the total source flux; in the descending order of the flux at 100 TeV, they are the Cygnus Cocoon, 1LHAASO J1825$-$1337u, 1LHAASO J1908$+$0615u, 1LHAASO J2228$+$6100u, 1LHAASO J1843$-$0335u, and 1LHAASO J1825$-$1256u. These six sources account for $\simeq 50\%$ of the total source flux at 100 TeV. The comparison shows that the gamma-ray flux of the resolved sources could significantly contribute to the sub-PeV Galactic gamma-ray emission, for example, $\simeq 66\%$ at 120 TeV.

However, the above source contribution must be regarded as a conservative upper limit in the sub-PeV energy range, as pointed out by \cite{Kato_2024_2}. In fact, the source contribution reaches $\simeq 85\%$ and $\simeq 110\%$ of the not-masked LHAASO flux at 500 TeV and 800 TeV, respectively, the latter of which is impossible. The former is even at odds with the observational fact that none of the 23 Tibet GDE events detected above $398\, {\rm TeV}$ in $22^{\circ}<l<225^{\circ}$ and $|b|<10^{\circ}$, except for the four events coming from the Cygnus Cocoon region, overlap with the sub-PeV LHAASO sources \citep{Kato_2024}. The discrepancy indicates that a simple extrapolation of the best-fit power-law spectra of the sub-PeV LHAASO sources well beyond 100 TeV is not an appropriate assumption to estimate the resolved-source contribution.

Indeed, spectral studies of several gamma-ray sources performed by LHAASO show a significant softening in the energy spectra in the several tens of TeV energy range \citep{LHAASO100TeV, Cao_2021_LHAASOJ0341+5258, LHAASO_W51, LHAASO_W43}. A large fraction of the first LHAASO catalog sources detected both by WCDA, which covers the energy range between $1\, {\rm TeV}$ and $25\, {\rm TeV}$, and KM2A, which operates above $25\, {\rm TeV}$, also show a clear sign of softening in their energy spectra from the measurements of the spectral indices in the corresponding energy ranges \citep{1LHAASOCatalog}. Therefore, the current gamma-ray observations imply that many of the sub-PeV LHAASO sources have a spectral cutoff or softening in the several tens of TeV energy range, and the total source flux estimated above should be regarded as a conservative upper limit in the sub-PeV energy range. Detailed spectral studies of individual gamma-ray sources would significantly reduce the fraction of the resolved-source contribution to the sub-PeV Galactic gamma-ray emission.

Furthermore, the GDE flux that contaminates the gamma-ray emissions of the sub-PeV LHAASO sources should also be revised. \cite{1LHAASOCatalog} estimate GDE contamination of the fluxes of the first LHAASO catalog sources in $|b| < 5^{\circ}$ by scaling their GDE measurement in $5^{\circ} < |b| < 10^{\circ}$ with the gas column density map created with the Planck dust opacity map \citep{Planck_2014, Planck_2016}. Their GDE measurement in $5^{\circ} < |b| < 10^{\circ}$ is well described by the model assuming a uniform cosmic-ray (CR) energy density and the gas distribution from the Plank data \citep{1LHAASOCatalog}, while the measurement in $|b| < 5^{\circ}$ shows an excess over the model as found by previous studies \citep{PhysRevLett.131.151001, LHAASO_diffuse_WCD_2024}. Therefore, the assumption of a uniform CR energy density may not be appropriate, which is implied in the GeV to TeV energy ranges \citep{PhysRevD.91.083012, Gaggero_2015, Acero_2016, PhysRevD.93.123007, PhysRevLett.119.031101}, and GDE contamination of the source fluxes should be higher. The increase in GDE contamination would reduce the intrinsic source fluxes and thus the resolved-source contribution to the total Galactic gamma-ray flux. Let us note that the above accurate estimate of the source contribution with respect to the total sub-PeV Galactic gamma-ray emission also allows us to study a potential space dependence of the CR spectrum covering the PeV energy range \citep{PhysRevD.110.103035}, the diffuse gamma-ray flux produced by the extragalactic CRs interacting with the Galactic interstellar medium, and sub-PeV gamma-ray emission from the decay of heavy dark matter in the Galaxy \citep{PhysRevLett.64.615, PhysRevLett.129.261103}.

The systematic uncertainty in the flux normalization in the LHAASO KM2A measurement should be studied further. Their flux systematic uncertainty of $7\%$ \citep{Aharonian_2021, 1LHAASOCatalog} only takes into account the time variation of the event rate due to that of the atmospheric density profile and does not include the uncertainty coming from the absolute energy-scale uncertainty. The latter could produce a particularly large flux systematic uncertainty; for example, in the case of Tibet AS$\gamma$, their absolute energy-scale uncertainty of $12\%$ estimated from the observation of the CR Moon's shadow \citep{Amenomori_et_al_2009} results in the flux systematic uncertainty of $+40\%\, -32\%$ under the gamma-ray differential spectral index of $-3$. The ARGO-YBJ experiment also estimates their absolute energy-scale uncertainty as $13\%$ \citep{PhysRevD.84.022003}. The accurate estimate of the absolute energy-scale uncertainty is important to constrain the absolute value of the sub-PeV Galactic gamma-ray flux and conclude the fractions of the GDE and source contributions.

The potential dominance of GDE in the sub-PeV energy range suggested above has some interesting physical implications. The EGRET and Fermi-LAT measurements have shown that GDE dominates in the GeV energy range \citep{Hunter_1997, Ackermann_2012}. They also found an excess emission in their GDE measurements against model predictions based on the distributions of the observed interstellar medium (ISM) and the local CR spectrum. One possible interpretation of the excess is a spatial dependence of the CR spectrum due to the anomalous CR diffusion caused by non-uniform magnetic turbulence in the ISM, which is supported by some gamma-ray observations and theoretical studies \citep{ERLYKIN201370, Acero_2016, PhysRevD.91.083012, Gaggero_2015, PhysRevD.93.123007, Luque_2023, Luque_2025}. The same physical mechanism may also apply to sub-PeV GDE to explain its potential dominance.

Another implication is that the sub-PeV gamma-ray emission from the Cygnus Cocoon detected by \cite{LHAASOCOLLABORATION2024449} has a significant GDE contamination. The authors claim that GDE only has a minor contamination to the total sub-PeV gamma-ray emission from the Cygnus Cocoon. However, their estimate of GDE contamination is based on the measurement by LHAASO \citep{PhysRevLett.131.151001} which masks a large fraction of the sky regions along the Galactic Plane to remove contributions from gamma-ray sources. \cite{Kato_2024_2} point out that the true GDE flux should thus be much higher, at least by a factor of $\sim 3$, compared to the flux measured by LHAASO in the inner Galactic Plane. The implication could thus significantly change the intrinsic gamma-ray flux of the Cygnus Cocoon, providing a key to accurately understanding the mechanism and energetics of CR acceleration in the source.

In conclusion, the Tibet GDE flux and the not-masked LHAASO flux are consistent, within the statistical and systematic uncertainties, for the inner Galactic Plane region in the sub-PeV energy range. The result suggests that the total sub-PeV Galactic gamma-ray emission is dominated by GDE. On the other hand, resolved sub-PeV gamma-ray sources (the sub-PeV LHAASO sources) could significantly contribute to the sub-PeV Galactic gamma-ray emission ($\approx 60\%$). However, the estimate must be regarded as a conservative upper limit in the sub-PeV energy range. In fact, current gamma-ray observations imply that many of the sub-PeV LHAASO sources have a spectral cutoff or softening in the several tens of TeV energy range, and the resolved-source contribution to the total sub-PeV Galactic gamma-ray emission should be much lower than the above estimate. More detailed discussion about the origin of the sub-PeV Galactic gamma-ray emission has need of spectral studies of individual resolved gamma-ray sources and an accurate estimate of GDE contamination of source fluxes.

\begin{acknowledgments}
  S. Kato, R. Alves Batista and C. Prévotat acknowledge support from the Agence Nationale de la Recherche (ANR), project ANR-23-CPJ1-0103-01. This work is supported in part by Grants-in-Aid for Scientific Research from the Japan Society for the Promotion of Science in Japan, the joint research program of the Institute for Cosmic Ray Research (ICRR), the University of Tokyo, and the use of the computer system of ICRR. This work is also supported by the National Natural Science Foundation of China under Grants No. 12227804, and the Key Laboratory of Particle Astrophysics, Institute of High Energy Physics, CAS.
\end{acknowledgments}

\end{document}